\documentclass[aps,reprint,onecolumn]{revtex4-1}

\usepackage{graphicx}
\usepackage{dcolumn}
\usepackage{amsfonts}
\usepackage{amsmath}
\usepackage{fancyhdr}

\begin{document}

\title{Multidimensional Coherent Photocurrent Spectroscopy of a Semiconductor Nanostructure}

\author{Ga\"el Nardin$^{1,4}$\footnote{gael.nardin@jila.colorado.edu}, Travis M. Autry$^{1,2,4}$, Kevin L. Silverman$^{3}$, Steven T. Cundiff$^{1,2}$}

\affiliation{$^1$ JILA, University of Colorado \& National Institute of Standards and Technology, Boulder, CO 80309-0440\\
$^2$ Department of Physics, University of Colorado, Boulder CO 80309-0390, USA\\
$^3$ National Institute of Standards and Technology, Boulder CO 80305-3328, USA\\ 
$^4$ These authors contributed equally to this work.}




\begin{abstract}
Multidimensional Coherent Optical Photocurrent Spectroscopy (MD-COPS) is implemented using unstabilized interferometers. Photocurrent from a semiconductor sample is generated using a sequence of four excitation pulses in a collinear geometry. Each pulse is tagged with a unique radio frequency through acousto-optical modulation ; the Four-Wave Mixing (FWM) signal is then selected in the frequency domain. The interference of an auxiliary continuous wave laser, which is sent through the same interferometers as the excitation pulses, is used to synthesize reference frequencies for lock-in detection of the photocurrent FWM signal. This scheme enables the partial compensation of mechanical fluctuations in the setup, achieving sufficient phase stability without the need for active stabilization. The method intrinsically provides both the real and imaginary parts of the FWM signal as a function of inter-pulse delays. This signal is subsequently Fourier transformed to create a multi-dimensional spectrum. Measurements made on the excitonic resonance in a double InGaAs quantum well embedded in a p-i-n diode demonstrate the technique.
\end{abstract}

\maketitle


%
%


\section{Introduction}
Multi-Dimensional Coherent Spectroscopy (MDCS) is an ultra-fast spectroscopy technique that unfolds spectral information onto several dimensions to correlate absorption, mixing, and emission frequencies. Originally developed for nuclear magnetic resonance experiments \cite{ernst_principles_1987}, MDCS was brought to the optical domain \cite{jonas_two-dimensional_2003}. In addition to molecules, this powerful technique has been very successfully applied to semiconductor heterostructures such as quantum wells \cite{li_many-body_2006,singh_anisotropic_2013}, or atomic vapors \cite{dai_two-dimensional_2010,li_unraveling_2013}. MDCS enables the separation of homogeneous and inhomogeneous line widths of a resonance \cite{singh_anisotropic_2013,siemens_resonance_2010,moody_exciton-exciton_2011}, the characterization of single and two-quantum coherences \cite{stone_two-quantum_2009,karaiskaj_two-quantum_2010}, and the study of the role played by many-body interactions \cite{li_many-body_2006}. It is also an ideal tool to study coupling between transitions, such as excitons confined in separated QWs \cite{davis_three-dimensional_2011,nardin_coherent_2013_arXiv} or Quantum Dots (QDs) \cite{moody_exciton-exciton_2011,kasprzak_coherent_2011,albert_microcavity_2013}. However, conventional MDCS techniques rely on the wave-vector selection of the FWM signal based on phase matching. This selection only applies in 2D or 3D systems such as atomic vapors, semiconductor QWs, or dense ensembles of nano-objects \cite{moody_exciton-exciton_2011}. Thus, conventional techniques cannot be applied to sub-wavelength structures where the translational symmetry is broken, such as individual (or small ensembles of) QDs, or other single nano-objects. In this case, radiation by point sources prevents the formation of a well defined FWM beam in the phase matched direction. So far, only a sophisticated heterodyne mixing technique has been demonstrated to provide access to FWM of single QDs \cite{kasprzak_coherent_2011}, relying on assumptions made about the phase evolution of a reference resonance in order to phase individual one-dimensional spectra.

Here we present a MDCS technique that enables measurement of FWM signals from nanostructures without relying on wave-vector selection of the signal. The approach measures the FWM signal via photocurrent readout and is suitable for any sample with which electrical contacts can be made, including single nano-objects. Advantages of the method include the intrinsic measurement of the FWM amplitude and phase, and a natural reduction of the impact of mechanical fluctuations, allowing us to achieve sufficient phase stability without active stabilization. As a proof of principle, 2D photocurrent spectra of excitons confined in a double InGaAs QW are presented.

\section{Experiment}

MDCS is an extension of FWM experiments. In conventional MDCS a sequence of three pulses is sent to the sample, and the radiated FWM is heterodyned with a local oscillator for phase resolution and detected by a spectrometer. Inter-pulse delays are precisely stepped,  and the data is then Fourier transformed with respect to these delays to obtain multi-dimensional spectra. This scheme has been implemented in various geometries. A common one is the pump-probe geometry, where the phase-locked delay between first and second pulses is achieved using pulse-shaping methods, and the signal is self-heterodyned with the transmitted probe serving as the local oscillator \cite{grumstrup_facile_2007}. The drawbacks of the pump-probe method are that only the real part (not the amplitude) of the correlation spectrum can be obtained, and delays are limited to about 10 ps (shorter than coherence and population times in semiconductor materials). Another implementation of conventional MDCS is the box geometry \cite{bristow_versatile_2009}. This geometry enables separation of rephasing and non-rephasing contributions \cite{li_many-body_2006}, the characterization of two-quantum coherences \cite{stone_two-quantum_2009,karaiskaj_two-quantum_2010}, and can provide longer inter-pulse delays if active stabilization is used \cite{bristow_versatile_2009}. However, MDCS in the box geometry requires additional procedures to obtain real and imaginary parts of the multidimensional spectra \cite{zhang_optical_2005,bristow_all-optical_2008}. Moreover, as mentioned above, non-collinear geometries cannot be used to study single or few zero-dimensional objects.

\begin{figure}[bp]
\centering
\includegraphics[width=12cm]{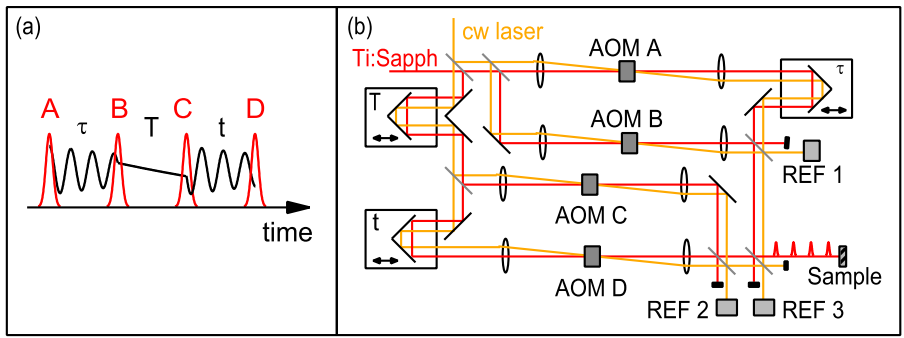}
\caption{(a) Pulse sequence used in the experiment. The four pulses form a collinear sequence. In a simple picture, the black line shows the first order polarization, second order population, and third order polarization generated by pulses A, B, and C, respectively. (b) Scheme of the experimental setup (described in text). Note: the cw laser is actually vertically offset, but shown horizontally offset in the figure for clarity.}
\label{Figure_setup}
\end{figure}

MDCS using collinear geometry has been demonstrated, based either on phase cycling algorithms \cite{tian_femtosecond_2003,aeschlimann_coherent_2011} or selection of the FWM signal in the frequency domain \cite{tekavec_fluorescence-detected_2007}. In the present work, we use a collinear geometry where four pulses are sent to the sample. Pulses are denoted A, B, C, and D (in the chronological order at which they hit the sample) and the corresponding inter-pulse delays are denoted $\tau$, $T$ and $t$. (Fig. \ref{Figure_setup} (a)). The role of the fourth pulse is to convert the third order polarization into a fourth order population that we detect in the form of a photocurrent signal. The FWM signal is isolated in the frequency domain, using an approach similar to that developed by Tekavec \textit{et al}, where the fourth order population generated in an atomic vapor was detected in the form of a fluorescence signal \cite{tekavec_fluorescence-detected_2007}. The four-pulse sequence is generated by sending the beam of a mode-locked Ti:Sapph oscillator (Coherent MIRA - 76 MHz repetition rate \cite{NIST_disclaimer}) into a set of two Mach-Zehnder interferometers nested within a larger Mach-Zehnder interferometer (Fig. \ref{Figure_setup} (b)). Three translation stages are used to control inter-pulse delays $\tau$, $T$ and $t$. Each interferometer arm contains an Acoustic Optical Modulator (AOM) (Isomet 1205c-1) driven by a phase-locked direct digital synthesizer (Analog Devices AD9959).  In each arm the 0$^{th}$ diffraction order of the AOM is spatially blocked and the 1$^{st}$ order beam is used. Each AOM tags the beams with a unique radio frequency $\omega_i$, with $i=\{A,B,C,D\}$. The resulting non-linear signals oscillate at radio frequencies given by sums and differences of the AOM frequencies.

\begin{figure}[bp]
\centering
\includegraphics[width=12cm]{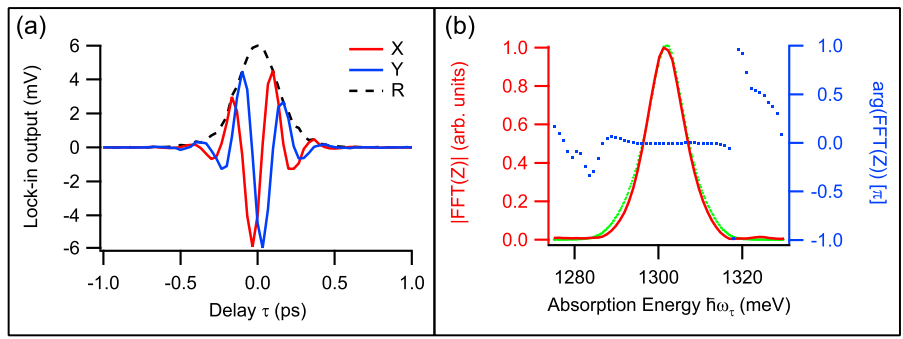}
\caption{(a) Result of a pulse-pulse correlation between pulses A and B, recorded on a broadband detector. $X$ and $Y$ are the in-phase and in-quadrature outputs of the lock-in amplifier, corresponding to the real and imaginary parts of the two-wave mixing signal $Z=X+iY$. $R=\sqrt{X^2+Y^2}$ is the field amplitude. $X$ and $Y$ oscillate with $\tau$ at the reduced frequency $\nu*$ $\sim$ 3.5 THz (see text). (b) Fast Fourier transform (FFT) of the complex signal. As the detector is broadband, $|FFT(Z)|$ (red curve) provides the power spectrum of the laser. The spectral phase is given by $arg(FFT(Z))$ (blue dots). A normalized spectrum of the excitation laser, recorded with an Ocean Optics USB 4000 spectrometer \cite{NIST_disclaimer}, is plotted for comparison (green circles).}
\label{Figure_autocorr}
\end{figure}

As a simple example, the photocurrent signal obtained from a combination of pulses A and B oscillates at the difference frequency (beat note) $\omega_{AB}=\omega_A-\omega_B$. Dual phase lock-in detection is used to isolate and demodulate this signal. A continuous wave (cw) laser (external cavity diode laser, $\lambda$ = 941.5) runs through the same optics as the excitation laser, but is vertically offset from the path of the excitation laser. The beat note resulting from its interference, recorded by reference photodetectors (REF1 for arms A and B on Fig. 1 (b)) serves as a reference for lock-in detection of the signal. This detection scheme provides several advantages: (1) The bandpass filtering around the beat note frequency achieved by the lock-in ensures that detected signal results from a wave-mixing process where pulses A and B contribute once each. Other contributions (single pulse contributions, noise at mechanical vibration frequencies, dark current) are filtered out by the lock-in. (2) The phase of the signal is correlated to the phase of the excitation pulses, providing a direct, intrinsic access to the complex signal Z = X + iY from the dual phase lock-in amplifier output. (3) When stepping delay $\tau$, the signal phase at the lock-in output does not evolve at an optical frequency, but at a reduced frequency $\nu^*=\left|\nu_{sig}-\nu_{cw}\right|$ given by the difference between the signal and cw laser optical frequencies. This ``physical undersampling'' of the signal results in a smaller number of points required to sample the signal, and in an improvement of the signal-to-noise ratio: the impact of a mechanical fluctuation in the setup, corresponding to a delay fluctuation $\delta\tau$, scales as $\delta\tau\cdot\nu^*$ \cite{tekavec_fluorescence-detected_2007,tekavec_wave_2006}. In other words, the cw laser senses and partially compensates for phase noise resulting from optical path fluctuations occurring in the setup. As a consequence, it is important to choose a cw laser wavelength as close as possible to the signal wavelength. With this technique, large inter-pulse delays are achievable (limited only by the coherence length of the cw laser or, as in our case, by the length of the delay stages). This is particularly useful for the spectroscopy of semiconductor nanostructures such as QDs, where coherence times can be of the order of the nanosecond \cite{borri_ultralong_2001,birkedal_long_2001}. We show in Fig. \ref{Figure_autocorr} the measurement to determine zero delay between pulses A and B. For this, we substitute for the sample with a broadband commercial detector (Thorlabs DET10A Si detector \cite{NIST_disclaimer}). Stepping the delay $\tau$, we perform a field auto-correlation of the excitation pulse. The signal, oscillating in real time at the frequency $\omega_{AB}$, is demodulated by the lock-in detection scheme. Figure \ref{Figure_autocorr} (a) shows the in-phase and in-quadrature components of the signal, provided by the lock-in outputs X and Y, respectively, and the signal amplitude R. X and Y, shifted by 90$^\circ$, evolve with the stepping of $\tau$ at the reduced frequency $\nu*=\left|\nu_{sig}-\nu_{cw}\right|$ $\sim$ 3.5 THz. Since the phase and amplitude of the signal are known, a Fourier transform with respect to $\tau$ provides us with the one-dimensional power spectrum and spectral phase of the excitation laser (Fig. \ref{Figure_autocorr} (b)).

In a similar way, the four pulses can be used to generate a FWM signal, and the delays $\tau$, $T$ and $t$ can be stepped to record 2D or 3D data, which can be Fourier transformed with respect to each delay to produce multi-dimensional spectra. The most typical 2D spectrum in conventional MDCS is obtained by stepping delays $\tau$ and $t$ while keeping $T$ at a constant value. To produce the appropriate lock-in reference for the FWM signal, we detect the beat notes $\omega_{AB}$ and $\omega_{CD}$ on two separate photodetectors (REF1 and REF2 in Fig. \ref{Figure_setup} (b)). These beat frequencies are mixed digitally using a digital signal processor (DSP) (SigmaStudios ADAU1761Z \cite{NIST_disclaimer}). The mixing algorithm (in-quadrature mixing) is as follows: the beat notes are input into the DSP, where they independently undergo a Hilbert transform. Then we make use of the identity $\cos(\omega_{AB} \pm \omega_{CD}) = \cos(\omega_{AB})\cos(\omega_{CD}) \mp \sin(\omega_{AB})\sin(\omega_{CD})$. This provides reference frequencies $\omega_{\ominus}=\omega_{CD}-\omega_{AB}=-\omega_A+\omega_B+\omega_C-\omega_D$ and $\omega_{\oplus}=\omega_{CD}+\omega_{AB}=\omega_A-\omega_B+\omega_C-\omega_D$ for the lock-in detection of the photocurrent FWM signal. These frequencies are analogous to the phase-matched directions $\vec{k}_{FWM}=-\vec{k}_A+\vec{k}_B+\vec{k}_C$ and $\vec{k}_{FWM}=\vec{k}_A-\vec{k}_B+\vec{k}_C$ of non-collinear MDCS, corresponding to the so-called rephasing (or $S_I$) and non-rephasing (or $S_{II}$) pulse sequences, respectively. In the box geometry, $S_I$ and $S_{II}$ need to be recorded separately since they correspond to a different pulse sequence, while with frequency domain selection $S_I$ and $S_{II}$ can be recorded simultaneously on two separate lock-in amplifiers. It is also possible to detect two-quantum (or $S_{III}$) signals that oscillate at the radio frequency $\omega_{S_{III}}=\omega_A+\omega_B-\omega_C-\omega_D$. The reference frequency that allows demodulation of such a signal cannot be generated from reference detectors REF1 and REF2. It can be extracted from detector REF3 at the output of the nested Mach-Zehnder interferometer (see Fig. \ref{Figure_setup}), and separated out from other frequency components using appropriate filters in the DSP.

\section{Results}

\begin{figure}[bp]
\centering
\includegraphics[width=12cm]{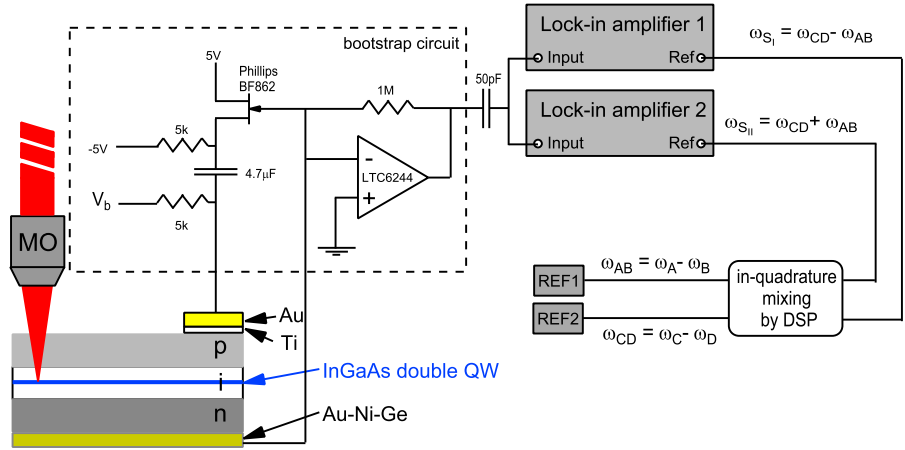}
\caption{Scheme of the electronic circuitry used to detect photocurrent from the sample and to generate lock-in references for rephasing ($S_I$) and non-rephasing ($S_{II}$) signals. MO: microscope objective. REF: reference photodetector.}
\label{Figure_electronics}
\end{figure}

We demonstrate the working principle of our setup on a double In$_{0.2}$Ga$_{0.8}$As/GaAs QW, embedded within the intrinsic region of a p-i-n diode. The double QW consists in a 4.8 nm thick QW and a 8 nm thick QW, separated by a barrier of 4 nm thickness. An Au-Ni-Ge bottom contact was deposited on the n-doped substrate. A top contact (5nm Ti and 200nm Au) was deposited on part of the sample surface. The sample was kept at a temperature of 15.5K in a cold finger liquid Helium cryostat. In order to deal with the sample capacitance, a ``bootstrap'' trans-impedance circuit \cite{graeme_photodiode_1996,LTC6244} was used to convert the photocurrent from the sample into a voltage that is read by the lock-in amplifiers (Stanford Research SRS830 \cite{NIST_disclaimer}) input (Fig. \ref{Figure_electronics}). 

The AOM frequencies that we use are $\omega_A$ = 80.109 MHz, $\omega_B$ = 80.104 MHz, $\omega_C$ = 80.019 MHz, $\omega_D$ = 80 MHz. The beat notes recorded by reference detectors REF1 and REF2 as a result of the cw laser interference are then $\omega_{AB}$ = 5 kHz and $\omega_{CD}$ = 19 kHz, respectively. As a result of in-quadrature mixing by the DSP, the reference frequencies provided to the lock-in amplifiers for the $S_I$ and $S_{II}$ FWM signal are $\omega_{S_I}$ = 14kHz and $\omega_{S_{II}}$ = 24kHz.

Let us underline that thanks to the ``physical undersampling'' phenomenon mentioned above, we do not need to implement active stabilization, a major advantage compared to conventional MDCS techniques \cite{bristow_versatile_2009}.

\begin{figure}[tbp]
\centering
\includegraphics[width=12cm]{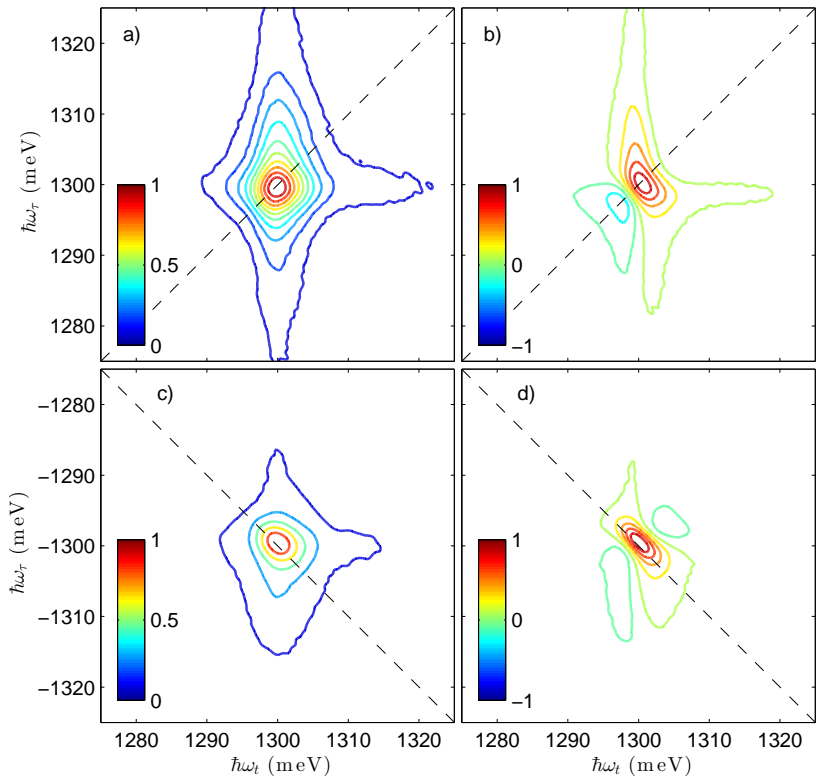}
\caption{(a) Absolute value, and (b) real part of the 2D spectrum recorded on the single InGaAs QW sample, using a non-rephasing pulse sequence ($S_{II}$). The spectra are plotted as a function of $\hbar\omega_\tau$ and $\hbar\omega_t$. (c) Absolute value, and (d) real part of the 2D spectrum recorded using a rephasing pulse sequence ($S_{I}$). The data to produce (a), (b), (c) and (d) were collected simultaneously.}
\label{Figure_2D}
\end{figure}

Figure \ref{Figure_2D} shows 2D spectra that were recorded from the sample. A forward bias $V_b$ = 0.5 V, for which we obtain the strongest FWM signal, was applied through the bootstrap circuit (see Fig. \ref{Figure_electronics}). The laser spectrum was the same as shown in Fig. \ref{Figure_autocorr} (b), exciting the lowest energy excitonic resonance of the double QW structure. The total excitation power (four pulses) was 250 $\mu W$. The pulse sequence was focused on the sample using a microscope lens (Nikon EPI ELWD CF Plan 20x, NA = 0.4 \cite{NIST_disclaimer}), providing an excitation spot of $\sim$5 $\mu$m diameter. While delays $\tau$ and $t$ were stepped, delay $T$ was kept at 200 fs. A fast Fourier transform with respect to delays $\tau$ and $t$ provides 2D spectra as a function of $\hbar\omega_\tau$ and $\hbar\omega_t$. Non-rephasing ($S_{II}$ -  Fig.\ref{Figure_2D} (a)-(b)) and rephasing (($S_{I}$ -  Fig.\ref{Figure_2D} (c)-(d)) spectra were obtained simultaneously from two separate lock-in amplifiers. 

The absolute value of $S_{II}$ and $S_{I}$ spectra (Fig.\ref{Figure_2D} (a) and (c)) show a peak on the diagonal corresponding to the lowest energy exciton of the double QW. In the $S_I$ spectrum the peak is slightly elongated along the diagonal, a sign of slight inhomogeneous broadening of the excitonic resonance. Thanks to the intrinsic phase resolution of the technique, real and imaginary parts of the data are directly obtained as well. To determine the absolute phase offset of the signal, we set the phase of the time domain data to be zero at zero $\tau$ and $t$ delays \cite{tekavec_fluorescence-detected_2007}: $arg\{Z(\tau=0,T=200fs,t=0)\}=0$. We show in Fig. \ref{Figure_2D} (b) and (d) the real parts of the $S_I$ and $S_{II}$ spectra, respectively, exhibiting a typical absorptive line shape.

\section{Conclusion}

We have implemented and demonstrated a robust, versatile platform for the MDCS of nanostructures. In a collinear geometry, the FWM signal is detected as a photocurrent and isolated in the frequency domain. This method provides an intrinsic phase resolution of the signal, and enables the recording of rephasing and non-rephasing FWM signals simultaneously. An auxiliary cw laser runs through the same optics as the excitation laser to partially compensate for mechanical fluctuations occurring in the setup, enabling optical MDCS without the need of active stabilization. Long inter-pulse delays can be achieved (only limited by the coherence length of the cw laser or the length of the delay stages), particularly adapted to the long coherence and population times of excitations in semiconductor materials. Demonstrated on an InGaAs double QW structure, the technique can be extended to the MDCS of single nano-objects, since it does not rely on the wave-vector selection of the FWM signal. Considering that photocurrent measurements have been reported in single QDs \cite{zrenner_coherent_2002,zecherle_ultrafast_2010}, carbon nanotubes \cite{bao_mapping_2012}, and nanowires \cite{wang_highly_2001,cao_engineering_2009,krogstrup_single-nanowire_2013}, the Multidimensional Coherent Optical Photocurrent Spectroscopy (MD-COPS) technique can be applied on all of these types of single nanostructures. Let us finally note that, while we have shown 2D rephasing and non-rephasing spectra, the MD-COPS setup can realize 3D spectra \cite{li_unraveling_2013}, as well as two-quantum 2D spectra, without any technical modification. 

We acknowledge Terry Brown, Andrej Grubisic and David Alchenberger for technical help and fruitful discussions, and Richard Mirin for the epitaxial growth of the QW sample. The work at JILA was primarily supported by the Chemical Sciences, Geosciences, and Energy Biosciences Division, Office of Basic Energy Science, Office of Science, U.S. Department of Energy under Award\# DE-FG02-02ER15346, as well NIST. G.N. acknowledges support by the Swiss National Science Foundation (SNSF).


\begin{thebibliography}{10}
\bibitem{NIST_disclaimer}
Mention of commercial products is for information only ; it does not imply NIST
  recommendation or endorsment, nor does it imply that the products mentioned
  are necessarily the best available for the purpose.
\end{thebibliography}
\end{document}